\documentclass[preprint,longnamesfirst]{emulateapj}
\usepackage{apjfonts,natbib}
\tighten

\bibpunct{(}{)}{;}{a}{}{,}

\slugcomment{The Astronomical Journal, accepted for the 2003 December issue}
\shortauthors{BAUER ET AL.}
\shorttitle{X-RAY PROPERTIES OF IC~342}

\begin{document}

\title{The X-ray Properties of the Nearby Star-Forming Galaxy IC~342: The {\it XMM-Newton} View}

\author{
F.~E.~Bauer,\altaffilmark{1}
W.~N.~Brandt,\altaffilmark{1}
and B.~Lehmer\altaffilmark{1}
}

\altaffiltext{1}{Department of Astronomy \& Astrophysics, 525 Davey Lab, 
The Pennsylvania State University, University Park, PA 16802.}

\begin{abstract}
We present results on the X-ray properties of IC~342 using a 10~ks
{\it XMM-Newton} observation. Thirty-five sources are detected
coincident with the disk of IC~342 (more than tripling the number
known), of which $\approx$~31 are likely to be intrinsic to
IC~342. This point-source population exhibits a diverse range of
spectral properties and has an X-ray luminosity function slope and
infrared luminosity comparable to that of starburst galaxies such as
M82 and the Antennae, while its relative lack of extended X-ray
emission is similar to the properties of quiescent spirals. Although
we find no evidence for short-term variability in any of the X-ray
sources, we do detect long-term variability between this observation
and the 1991 {\it ROSAT} and 1993/2000 {\it ASCA} observations for
five sources. Notably, the second most luminous source in IC~342 (X-2)
is found to have an absorption-corrected 0.5--10~keV luminosity of
$5.75\times10^{39}$~erg~s$^{-1}$ and a spectrum best-fit by an
absorbed multi-color accretion-disk model with $T_{\rm
in}=2.17^{+0.25}_{-0.12}$~keV and $N_{\rm
H}=(15.3^{+1.9}_{-1.8})\times10^{21}$~cm$^{-2}$. This is the lowest
luminosity state observed for X-2 to date, although the slope of the
spectrum is intermediate between the previously observed low/hard and
high/soft states. IC~342 X-1, on the other hand, is found to be in an
identical state to that observed in 2000 with {\it ASCA}. Assuming X-1
is in an anomalous very high (VH) state, then either (1) X-1 has
remained in this state between 2000 and 2002, and is therefore the
longest duration VH-state binary ever observed, or (2) it was simply
caught in a VH state by chance in both the 2000 {\it ASCA} and 2002
{\it XMM-Newton} observations. We have also confirmed the {\it ROSAT}
HRI result that the nucleus of IC~342 is made up of both point-like
and extended emission, with the extended emission contributing
$\approx$55\% and $\approx$35\% in the 0.3--2.0~keV and 2.0--10.0~keV
bands, respectively. The spectrum of the nucleus is best fit by an
absorbed two-component model consisting of a thermal plasma with a
temperature of $kT\approx0.3$ and a power law with a photon index of
$\Gamma\approx2.53$. The relative fluxes of the two spectral
components suggest that the nucleus is complex, with a soft extended
component contributing approximately half of the total luminosity.
\end{abstract}

\keywords{galaxies: active --- 
galaxies: general --- 
X-rays: binaries --- 
X-rays: galaxies
}

\section{Introduction}\label{intro}

IC~342 is a nearby, nearly face-on ($i\sim20^{\circ}$) Scd galaxy with
spiral arms undergoing moderate star formation and an intense
starburst core. It has an absolute visual magnitude of
$M_{V}\approx-21.3$, an optical $D_{25}$ diameter of
$\approx$~29.7$\arcmin$, and neutral hydrogen and total dynamical
masses of $M_{\rm HI}=2.1\times10^{9}$~$M_{\odot}$ and $M_{\rm
tot}=1.1\times10^{11}$~$M_{\odot}$, respectively \citep{Tully1988,
Crosthwaite2000}. Although the Galactic absorption column density
toward IC~342 is relatively large \citep[$N_{\rm
H}=3\times10^{21}$~cm$^{-2}$;][]{Stark1992}, it has been observed by
{\it Einstein}, {\it ROSAT}, and {\it ASCA}. {\it ROSAT} High
Resolution Imager (HRI) observations ($\theta_{\rm \small
FWHM}\approx$~5\arcsec) resolved the disk of the galaxy into a
collection of ten point sources above a 0.1--2.5~keV flux limit of
$\approx$~$3\times10^{-14}$~erg~cm$^{-2}$~s$^{-1}$
\citep[or an absorption-corrected luminosity limit of
$\approx$~$2\times10^{37}$~erg~s$^{-1}$;][hereafter
BCT93]{Bregman1993}.\footnote{We adopt a distance of $3.3\pm0.3$~Mpc
to IC~342, based on the period-luminosity relation of 20 Cepheid
variables \citep{Saha2002}. Results from other papers have been
corrected to this distance.} {\it ROSAT} and {\it ASCA} measurements
indicate that the bulk of the X-ray emission arises from four
ultraluminous X-ray sources (ULXs, \hbox{$L_{\rm
X}\approx$~$10^{39}$--$10^{40}$~erg~s$^{-1}$}), one of which is
coincident with the nucleus \citetext{BCT93; \citealp{Okada1998},
hereafter O98; \citealp{Kubota2001}, hereafter K01}. Two {\it ASCA}
observations (38~ks in 1993, 276~ks in 2000) have shown the ULXs to
vary both in intensity and spectral shape \citetext{O98; K01;
\citealp{Sugiho2001}}, suggesting that they are single objects with
masses of \hbox{$\approx$~10--100~$M_{\odot}$} (if radiating
isotropically below their Eddington limits). The nucleus, however, has
remained constant and appears to be slightly extended (BCT93). Since
this extended emission encompasses the surface-brightness enhancements
seen at optical-to-radio wavelengths (\hbox{$\approx$~10--15\arcsec})
that are attributed to a high rate of ongoing star formation, BCT93
speculated that the emission might be from a hot gas bubble.

Here we report results from a 10~ks {\it XMM-Newton} observation of
IC~342. This observation provides the highest resolution and most
sensitive hard X-ray imaging of IC~342 published to date, as well as a
field of view which covers the entire extent of the galaxy. We
describe the observation and reduction of the X-ray data in
$\S$\ref{data}, provide the basic X-ray properties of the detected
sources in the context of previous results in $\S$\ref{properties},
and summarize our findings in $\S$\ref{conclusions}.

\section{XMM--Newton Observation and Data Analysis}\label{data}

IC~342 was observed on 2001 February 11 for $\approx$10~ks with the
PN--CCD camera \citep{Struder2001} and the two MOS--CCD cameras
\citep{Turner2001} using the medium filter. The nucleus of IC~342 was
placed at the aimpoint, allowing the entire optical extent of the
galaxy to be imaged. The processing, screening, and analysis of the
data were performed using the standard tools from {\tt SAS} (v.5.3.3)
and {\tt CIAO} (v2.3), as well as custom {\tt IDL} software. The raw
MOS and PN data were initially processed using the standard {\it
epchain} and {\it emchain} pipeline scripts. Time intervals
contaminated by soft-proton flares were identified using the
background light curve in the 10--13~keV band. Approximately 200~s of
exposure was excluded at the end of the observation due to flaring;
the background for the rest of the observation remained relatively
constant at $\la0.2$~counts~s$^{-1}$. Our final exposures were 9488~s,
9507~s, and 4888~s for the MOS1, MOS2, and PN detectors,
respectively. We selected only good event patterns: $\le$12 for MOS
imaging and spectroscopy, $\le$12 for PN imaging, and $\le$4 for PN
spectroscopy.

\begin{figure}
\vspace{-0.0in}
\centerline{
\includegraphics[height=8.5cm]{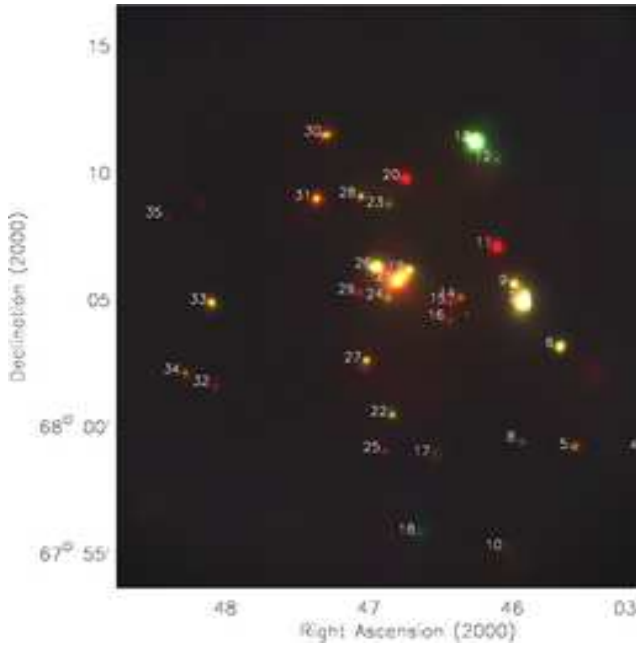}
}
\vspace{-0.1cm} \figcaption[IC342.eps]{ An {\it XMM-Newton}
``false-color'' $25\arcmin \times 25\arcmin$ image of IC~342 with red,
green, and blue representing 0.3--2.0~keV, 2.0--5.0~keV, and
5.0--10~keV emission, respectively. Prior to combination, each image
was smoothed with an adaptive-kernel algorithm
\citep{Ebeling2003}. Little diffuse emission is observed. In addition
to a bright nuclear component (source 21), 34 off-nuclear point
sources are detected. The sources have been labeled according to their
source number in Table~\ref{tab:sources}. 
\label{fig:smooth}}
\vspace{-0.1in} 
\end{figure}

Source detection was initially performed on the MOS and PN images in
the 0.3--10~keV band using the standard {\tt SAS} {\it eboxdetect} and
{\it emldetect} algorithms. A likelihood value of 10 was imposed,
corresponding to a significance level of $\approx$3.6$\sigma$. Thirty
sources were detected, although we found that several apparent sources
--- including two with $\ga$100 MOS$+$PN counts --- were detected by
{\it eboxdetect} but rejected by {\it emldetect}.\footnote{These
sources are noted in Table~\ref{tab:sources}.} As a cross-check, we
compared the sources found using {\it eboxdetect} with those found
using the {\tt CIAO} {\it wavdetect} algorithm \citep{Freeman2002},
run on the merged MOS$+$PN image with a significance threshold of
$1\times10^{-7}$. We found good agreement between the algorithms, and
we have chosen the 35 sources common to both methods as our X-ray
point-source sample. Table~\ref{tab:sources} lists the basic X-ray
properties of these sources, while Figure~\ref{fig:smooth} indicates
their position in the X-ray image. Background-subtracted counts were
extracted using circular apertures ranging in radius from
\hbox{$\approx$17--68\arcsec}, depending on off-axis angle and the relative
brightness of the source compared to the local background. Local
backgrounds were determined after removal of the 35 point sources. Our
background-subtracted detection limit is $\approx 25$ MOS$+$PN counts,
corresponding to an absorbed 0.3--10~keV flux of $\approx
(1$--$2)\times10^{-14}$~erg~cm$^{-2}$~s$^{-1}$.

\begin{figure}
\vspace{-0.1in}
\centerline{
\includegraphics[height=8.5cm]{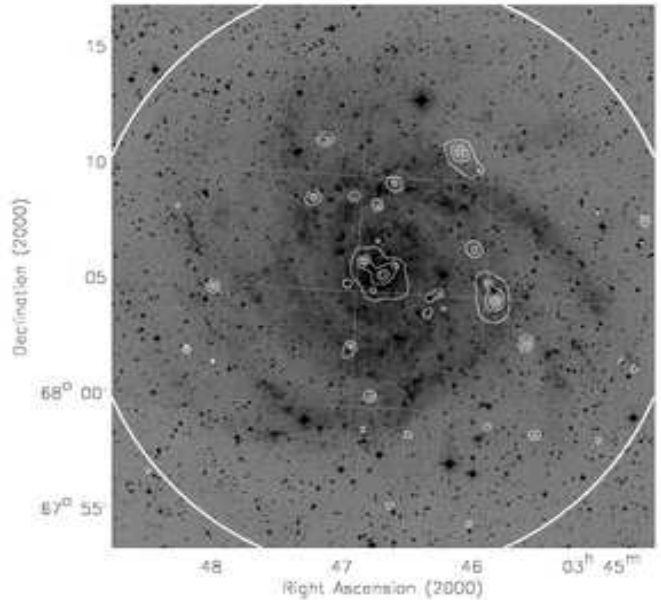}
}
\vspace{-0.1cm} 
\figcaption[IC342.eps]{
A $25\arcmin \times 25\arcmin$ Digitized Sky Survey $B_{J}$-band
image with {\it XMM-Newton} contours superimposed. Contours are
calculated from the combined MOS image for the 0.3--10~keV band. The
lowest contour indicates 0.33~counts~pixel$^{-1}$ ($\approx3\times$
the average background value), with each subsequent contour increasing
by a factor of four. All sources lie within the white $12\arcmin$
radius circle shown.
\label{fig:overlay}}
\vspace{0.2cm} 
\end{figure}

We have more than tripled the number of known X-ray sources spatially
coincident with IC~342, compared to the {\it ROSAT} HRI observation
reported by BCT93 of similar soft-band depth, thanks to the
high-energy response of {\it XMM-Newton}. Figure~\ref{fig:overlay}
shows X-ray contours overlaid on an optical image of IC~342,
indicating that the majority of the X-ray sources coincide with the
spiral arms of the galaxy.

The nominal astrometric accuracy for {\it XMM-Newton} is
\hbox{$\approx4$\arcsec} \citep[e.g.,][]{Jansen2001}. We have improved
upon this somewhat by matching X-ray sources to optical sources from
the $r$-band image of \citet{Saha2002} and the United States Naval
Observatory B1.0 catalog \citep[USNOB1;][]{Monet2003}.\footnote{The
$r$-band image was astrometrically registered to an accuracy of
$\approx$0\farcs1 RMS using 216 stars from the Tycho-2 and Guide Star
v2.2 catalogs.} X-ray matching to optical sources was performed using
a 4$\arcsec$ matching radius, with 13 sources having potential optical
counterparts: the nucleus of IC~342, four spatially extended sources
(potential H{\sc ii} regions), five faint point sources with
$R\sim21$--25 within the spiral arms of IC~342, and three bright
foreground stars with $R\sim11$--13. Excluding the five faint point
sources that potentially could be chance alignments, we find average
{\it XMM-Newton} Right Ascension and Declination offsets of
$\approx$~$-3$\farcs4 and 0\farcs4, respectively, resulting in a
registration accuracy of $\approx$1\farcs3~rms; positions have been
corrected for these offsets throughout. These offsets do not change
significantly (only by 0\farcs3) if we align the X-ray and optical
images using only the three bright stars and the nucleus (i.e., the
most secure matches). Although we have limited statistics, we find no
obvious rotation of the X-ray image to within our registration errors.

\tabletypesize{\scriptsize}
\begin{deluxetable*}{rllrrrrrrrcl}
\tablewidth{0pt}
\tablecaption{X-ray Sources Spatially Coincident with IC~342\label{tab:sources}} 
\tablehead{
\colhead{(1)} & 
\colhead{(2)} & 
\colhead{(3)} & 
\colhead{(4)} & 
\colhead{(5)} & 
\colhead{(6)} & 
\colhead{(7)} & 
\colhead{(8)} & 
\colhead{(9)} & 
\colhead{(10)} & 
\colhead{(11)} &
\colhead{(12)} \\
\colhead{ID} & 
\colhead{Source Name} & 
\colhead{Other Names} & 
\colhead{Off-Axis} & 
\colhead{Counts} & 
\colhead{BR} & 
\colhead{$F_{\rm X}$} & 
\colhead{$L_{\rm X}$} & 
\colhead{$R$} & 
\colhead{log($F_{\rm X}/F_{\rm O}$)} & 
\colhead{HRI $F_{\rm X}$} & 
\colhead{Optical ID} \\
\colhead{} & 
\colhead{XMMU~J} & 
\colhead{} & 
\colhead{$\arcmin$} & 
\colhead{} & 
\colhead{} & 
\colhead{} & 
\colhead{} & 
\colhead{} & 
\colhead{} & 
\colhead{} & 
\colhead{}
}
\startdata
 1        & 034448.3$+$680843           &             & 11.6 & $ 107.3^{+13.7}_{-11.8}$ & $0.54^{+0.16}_{-0.14}$ &     14.20 &     2.43 &      $\ga20$ & $\ga 0.65$ & $<$3.74 & \\
 2{\rlap*}& 034449.4$+$680738           &             & 11.3 & $  22.8^{ +9.3}_{ -7.4}$ & $<$ 0.98               &      0.21 & Galactic &      13.0    & $-3.98$    & $<$3.17 & Star (1\farcs5) \\
 3        & 034449.7$+$680214           &             & 11.7 & $  39.8^{+11.2}_{ -9.3}$ & $1.29^{+0.75}_{-0.61}$ &      3.78 &     0.65 &      $\ga20$ & $\ga 0.08$ & $<$2.46 & \\
 4        & 034505.1$+$675904           &             & 11.8 & $  28.1^{ +9.3}_{ -7.4}$ & $1.43^{+0.99}_{-0.78}$ &      0.92 &     0.16 &      $\ga20$ & $\ga-0.54$ & $<$2.74 & \\
 5        & 034534.7$+$675908$^\dagger$ & R1          &  9.6 & $  87.1^{+12.9}_{-11.0}$ & $0.74^{+0.23}_{-0.20}$ &      9.71 &     1.66 &      $\ga20$ & $\ga 0.49$ & $<$4.06 & \\
 6        & 034540.2$+$680308$^\dagger$ &             &  6.9 & $ 325.9^{+22.3}_{-20.4}$ & $1.22^{+0.17}_{-0.16}$ &     23.91 &     5.92 &      $\ga20$ & $\ga 0.88$ &    3.16 & \\
 7        & 034556.0$+$680455           & X-1, R3, A1 &  5.0 & $4132.8^{+68.3}_{-66.5}$ & $1.07^{+0.04}_{-0.03}$ &    214.54 &    50.25 &     20.1$^S$ & $1.96$     &   27.17 & Supernova Remnant \\
          &                             &             &      &                          &                        &           &          &              &            &         &\citep[0\farcs3;][]{Roberts2003} \\
 8        & 034556.8$+$675926           &             &  8.0 & $  54.1^{+11.2}_{ -9.3}$ & $0.89^{+0.38}_{-0.32}$ &      2.96 &     0.51 &      $\ga20$ & $\ga-0.03$ & $<$2.48 & \\
 9        & 034559.7$+$680536           &             &  4.6 & $ 324.4^{+20.9}_{-19.1}$ & $0.98^{+0.13}_{-0.12}$ &     20.04 &     4.69 &      $\ga20$ & $\ga 0.80$ & $<$4.91 & \\
10        & 034603.2$+$675510           &             & 11.4 & $  30.0^{ +9.1}_{ -7.1}$ & $0.56^{+0.43}_{-0.32}$ &      3.11 &     0.53 &      $\ga20$ & $\ga-0.01$ & $<$2.73 & \\
11{\rlap*}& 034606.8$+$680705           & R4          &  4.1 & $ 275.9^{+21.0}_{-19.2}$ & $0.04^{+0.04}_{-0.03}$ &      3.65 & Galactic &      12.0    & $-3.05$    &    4.07 & Star (0\farcs6) \\
12        & 034607.0$+$681029           &             &  6.1 & $  72.9^{+11.9}_{-10.0}$ & $1.70^{+0.57}_{-0.52}$ &      6.22 &     1.06 &     24.7$^S$ & $1.17$     & $<$2.15 & (0\farcs5)\\
13        & 034616.0$+$681115           & X-2, R5, A2 &  6.3 & $2215.9^{+51.0}_{-49.3}$ & $3.41^{+0.19}_{-0.19}$ &    208.66 &    50.02 &    $\ga24^S$ & $\ga 3.48$ &   12.56 & \\
14        & 034622.4$+$680505$^\dagger$ &             &  2.6 & $  55.1^{+11.4}_{ -9.5}$ & $0.38^{+0.20}_{-0.16}$ &      6.55 &     1.12 &     20.2$^S$ & $0.40$     & $<$4.17 & H{\sc ii} Region? (1\farcs3, also $R=21.8$, 1\farcs9) \\
15{\rlap*}& 034626.4$+$680453$^\dagger$ &             &  2.3 & $  37.0^{+10.8}_{ -8.9}$ & $<$ 0.52               &      0.34 & Galactic &     12.9$^S$ & $-3.81$    & $<$2.44 & Star (2\farcs3) \\
16        & 034627.2$+$680414$^\dagger$ &             &  2.5 & $  44.3^{+10.8}_{ -8.9}$ & $0.26^{+0.20}_{-0.15}$ &      2.39 &     0.41 & $\sim$21$^S$ & $\sim0.28$ & $<$3.13 & (0\farcs2; complex X-ray source \\
          &                             &             &      &                          &                        &           &          &              &            &         & blend with nearby star?) \\
17        & 034633.2$+$675852           &             &  7.0 & $  46.5^{+10.6}_{ -8.7}$ & $1.10^{+0.51}_{-0.43}$ &      4.15 &     0.71 &      $\ga20$ & $\ga 0.12$ & $<$3.71 & \\
18        & 034639.6$+$675548           &             & 10.0 & $  33.7^{ +9.4}_{ -7.4}$ & $>$ 2.92               &      8.40 &     1.43 &      $\ga20$ & $\ga 0.42$ & $<$3.32 & Hard X-ray source, possible bkgd. AGN? \\
19        & 034643.9$+$680609           & R6          &  0.6 & $ 402.4^{+22.8}_{-21.0}$ & $0.72^{+0.08}_{-0.08}$ &     14.87 &     3.51 &     21.8$^S$ & $1.53$     &    2.68 & (1\farcs0) \\
20{\rlap*}& 034646.0$+$680946           & R7          &  4.0 & $ 181.2^{+18.3}_{-16.4}$ & $<$ 0.15               &      4.07 & Galactic? &    21.7$^S$ & $0.84$     &   12.85 & Star? (0\farcs3) \\
21        & 034648.8$+$680546           & X-3, R8, A3 &  0.0 & $2233.1^{+50.2}_{-48.4}$ & $0.30^{+0.02}_{-0.02}$ &     54.16 &    15.05 & $\sim$10$^S$ & $\sim-2.7$ &   21.74 & Nucleus (0\farcs3) \\
22        & 034651.2$+$680027$^\dagger$ &             &  5.3 & $ 108.3^{+14.7}_{-12.9}$ & $1.86^{+0.56}_{-0.50}$ &      7.46 &     1.27 &      $\ga20$ & $\ga 0.37$ & $<$2.70 & \\
23        & 034652.8$+$680841           &             &  2.9 & $  56.7^{+13.8}_{-11.9}$ & $1.49^{+0.84}_{-0.65}$ &      3.44 &     0.59 &     21.9$^S$ & $0.80$     & $<$2.36 & H{\sc ii} Region? (0\farcs2) \\
24        & 034652.9$+$680504           &             &  0.8 & $ 106.8^{+13.7}_{-11.8}$ & $0.63^{+0.17}_{-0.15}$ &      6.93 &     1.18 &    $\ga22^S$ & $\ga 1.14$ & $<$3.41 & \\
25        & 034654.2$+$675902           &             &  6.8 & $  27.7^{ +8.5}_{ -6.5}$ & $<$ 0.66               &      3.99 &     0.68 &      $\ga20$ & $\ga 0.10$ & $<$2.64 & \\
26        & 034657.7$+$680616           & R9          &  1.0 & $1204.9^{+37.4}_{-35.7}$ & $0.79^{+0.05}_{-0.05}$ &     41.34 &     8.96 &    $\ga22^S$ & $\ga 2.02$ &   10.50 & \\
27        & 034702.2$+$680234           &             &  3.4 & $ 114.2^{+13.6}_{-11.7}$ & $0.94^{+0.22}_{-0.20}$ &      8.32 &     1.42 &      $\ga20$ & $\ga 0.45$ & $<$3.32 & \\
28        & 034704.5$+$680904           &             &  3.6 & $  89.5^{+12.8}_{-10.9}$ & $1.05^{+0.30}_{-0.27}$ &      6.39 &     1.09 &    $\ga24^S$ & $\ga 1.91$ & $<$2.50 & \\
29        & 034704.8$+$680515           &             &  1.6 & $  42.0^{+10.2}_{ -8.3}$ & $0.20^{+0.18}_{-0.13}$ &      2.30 &     0.39 &      18.5    & $-0.74$    & $<$3.54 & H{\sc ii} Region? (1\farcs3) \\
30        & 034719.2$+$681129           &             &  6.4 & $ 198.8^{+18.5}_{-16.7}$ & $0.28^{+0.08}_{-0.07}$ &      9.85 &     5.66 &     21.4$^S$ & $1.13$     & $<$3.07 & (1\farcs0) \\ 
31        & 034723.3$+$680857           &             &  4.5 & $ 267.2^{+20.1}_{-18.3}$ & $0.34^{+0.07}_{-0.06}$ &      6.90 &     7.71 &      $\ga20$ & $\ga 0.42$ & $<$2.73 & \\
32        & 034805.9$+$680138           &             &  8.3 & $  25.0^{ +9.3}_{ -7.3}$ & $<$ 0.77               &      2.38 &     0.41 &      $\ga20$ & $\ga-0.12$ & $<$3.13 & \\
33        & 034807.4$+$680451           & R10         &  7.4 & $ 202.8^{+18.3}_{-16.5}$ & $0.67^{+0.13}_{-0.12}$ &     10.05 &     2.81 &      $\ga20$ & $\ga 0.66$ &    4.70 & \\
34        & 034818.3$+$680205           &             &  9.1 & $  67.2^{+11.8}_{ -9.9}$ & $0.74^{+0.28}_{-0.24}$ &      7.62 &     1.30 &      $\ga20$ & $\ga 0.38$ & $<$2.74 & \\
35        & 034826.4$+$680815           &             &  9.4 & $  22.5^{ +9.6}_{ -7.6}$ & $0.55^{+0.60}_{-0.45}$ &      2.87 &     0.49 &      $\ga20$ & $\ga-0.04$ & $<$2.48 & \\
\enddata
\tablecomments{
Column 1: ID number. Sources denoted by a ``*'' are not thought to be intrinsic to IC~342.
Column 2: Source name given as XMMU~JHHMMSS.S$+$DDMMSS. Sources
denoted by a $\dagger$ were rejected by {\it emldetect} for no obvious
reason (see $\S$\ref{data}).
Column 3: Other names. R\# and A\# indicate the {\it ROSAT} and {\it
ASCA} source numbers assigned by BCT93 and O98, respectively.
Column 4: Off-axis angle in units of arcminutes.
Column 5: Background-subtracted 0.3--10~keV counts, summed over all of
the EPIC instruments (pn$+$MOS1$+$MOS2). Aperture photometry was
performed using extraction radii ranging in radius from
$\approx$17--68$\arcsec$, depending on off-axis angle and the relative
brightness of the source compared to the local background (see
$\S$\ref{data}). The errors for the source and background counts were
computed following the method of \citet{Gehrels1986} and were then
combined following the ``numerical method'' described in $\S$1.7.3 of
\citet{Lyons1991}.
Column 6: Band ratios, defined as the ratio of counts between the hard
and soft bands. The quoted band ratios have been corrected for
differential vignetting between the hard band and soft band using the
appropriate exposure maps. Errors for this quantity are calculated
following the ``numerical method'' described in $\S$1.7.3 of
\citet{Lyons1991}.
Column 7: Observed, aperture-corrected 0.5--10~keV fluxes in units
of 10$^{-14}$~erg~cm$^{-2}$~s$^{-1}$ from the best-fit models to the
{\it XMM-Newton} spectra (see Table~\ref{tab:spectra}). For faint
sources ($<$ 150 counts), fluxes were calculated assuming an average
power-law spectrum with $N_{\rm H} = 3.4\times10^{21}$~cm$^{-2}$ and a
photon index of $\Gamma=1.63$ as determined from the stacked spectrum of
the faint sources.
Column 8: Absorption-corrected 0.5--10~keV luminosities in units of
10$^{38}$~erg~s$^{-1}$ from the best-fit models to the {\it
XMM-Newton} spectra.
Column 9: $R$ magnitude as derived from the \citet{Saha2002} $r'$
image (indicated by an ``S'') or USNOB1. The magnitude lower limits
were estimated based on the faintest detectable sources within the
vicinity of the X-ray source.
Column 10: X-ray-to-optical flux ratio using columns 7 and 9 such that 
log($F_{\rm X}/F_{\rm O}$)$=$log($F_{\rm X}$)$+$5.5$+R/2.5$.
Column 11: Observed, aperture-corrected 0.5--2~keV fluxes in units of
10$^{-14}$~erg~cm$^{-2}$~s$^{-1}$ from the {\it ROSAT} HRI derived
using XIMAGE and PIMMS. The adopted spectral models were identical to
those assumed in Column 5.
Column 13: Possible optical ID and additional comments. Saha and
USNOB1 offsets are provided in parentheses for sources with optical
counterparts.}
\end{deluxetable*}

Because IC~342 has a large angular extent and lies at low Galactic
latitude ($b=10\fdg58$), the chance superposition of Galactic X-ray
sources and background AGN is a concern. Fortunately, IC~342 has a
Galactic longitude of $l=138\fdg17$, so our line-of-sight is
relatively far from the Galactic Center and therefore intersects only
a relatively small fraction of the Galaxy. Based on the numbers of
serendipitous sources detected in four archival observations within
15$^{\circ}$ of IC~342 (Proposal IDs 00011, 10946, 11220, 11229), and
from extrapolation of the medium-deep {\it XMM-Newton} $\log N$--$\log
S$ of \citet[][ i.e., $\approx$~60--100 sources per deg$^2$ down to a
limiting 0.5--10~keV flux of $\approx
1\times10^{-14}$~erg~cm$^{-2}$~s$^{-1}$]{Baldi2002}, we would expect
$\sim$~7--12 foreground or background sources within a 12$\arcmin$
radius of IC~342. Given that we are able to identify four of our 35
sources with foreground stars (sources 2, 11, 15, and 20), we suspect
that the majority of the remaining 31 X-ray sources are associated
with IC~342 based on their large X-ray-to-optical flux ratios; half
have log$(F_{\rm X}/F_{\rm O})\ga0.5$, and all but the nucleus have
log$(F_{\rm X}/F_{\rm O})\ga-0.5$.  A comparison of the
X-ray-to-optical flux ratios of these sources with Figure~1 of
\citet[][after accounting for average X-ray and optical band
differences]{Maccacaro1988} suggests these sources are inconsistent
with X-ray emission from all normal stars and even some AGN (the most
likely foreground and background contaminants, respectively), but they
are fully consistent with X-ray binaries in IC~342.  Additionally,
sources that are significantly brighter than our limiting X-ray flux,
and are therefore rarer, are even more likely to be associated with
IC~342. For instance, there is a $\la$ 0.1\% chance that any of the
four sources brighter than $\approx
5\times10^{-13}$~erg~cm$^{-2}$~s$^{-1}$ are foreground or background
X-ray sources.

Point-source spectra were extracted using the apertures described
above. Event PI values and photon energies were determined using the
latest gain files appropriate for the observation. The X-ray spectra
were analyzed using {\tt XSPEC} \citep{Arnaud1996}. Unless stated
otherwise, spectral parameter errors are for 90\% confidence assuming
one parameter of interest ($\Delta\chi^2=2.7$). The X-ray fluxes and
absorption-corrected luminosities for all 35 sources were calculated
from spectral fitting using {\tt XSPEC}. None of the sources was
affected by pile-up. For faint sources ($<$ 150 counts) thought to be
intrinsic to IC~342 (i.e., not obvious foreground stars), fluxes were
calculated assuming an average absorbed power-law spectrum with
$N_{\rm H} = (3.4^{+0.9}_{-0.7})\times10^{21}$~cm$^{-2}$ and a photon
index of $\Gamma=1.63^{+0.13}_{-0.14}$, as determined from the stacked
spectrum of the faint sources. Fluxes for the two faint foreground
stars (sources 2 and 15) were calculated assuming the spectrum of
source 11 (see Table~\ref{tab:spectra}). Note that if a faint source's
spectrum deviates substantially from these average values, then the
flux and absorption-corrected luminosity of the source may change as
well.

\section{IC~342 X-ray Properties}\label{properties}

The 31 point sources plausibly associated with IC~342 span over two
orders of magnitude in X-ray luminosity and display a diverse range of
spectral properties, as implied from the band ratios listed in
Table~\ref{tab:sources} ($BR \sim$ 0.1--3.4). The \hbox{0.5--10~keV}
X-ray luminosity function (XLF) of the sources is relatively flat down
to $L_{\rm X}\sim4\times10^{37}$~erg~s$^{-1}$ with a slope of
$0.5\pm0.1$, as determined from maximum likelihood fitting of the
differential XLF. At luminosities fainter than this limit,
incompleteness due to the sensitivity limit and absorption bias
against soft sources makes the XLF slope uncertain. 
Somewhat surprisingly, we detect little diffuse X-ray emission (only a
compact region around the nucleus; see $\S$\ref{X3_spatial}). In
estimating the amount of diffuse emission, we only used the combined
MOS1$+$MOS2 0.3--2.0~keV image for simplicity. We extracted a total of
$124\pm48$ counts within a 5$\arcmin$ radius aperture around the
nucleus, using a vignetting-corrected 5--10$\arcmin$ background
annulus and masking out point sources using their 99\%
encircled-energy radius apertures. Assuming this emission would have
an absorbed thermal plasma spectrum with a temperature of 0.5~keV
(1~keV) and column density of $N_{\rm H}=6\times10^{21}$~cm$^{-2}$
(see $\S$\ref{spectra}), our 3$\sigma$ upper limit corresponds to a
flux density of
$<8.4\times10^{-15}$~erg~cm$^{-2}$~s$^{-1}$~arcmin$^{-2}$
($<9.9\times10^{-16}$~erg~cm$^{-2}$~s$^{-1}$~arcmin$^{-2}$) and an
absorption-corrected 0.5--2.0~keV luminosity of $L_{\rm
X}<5.6\times10^{39}$~erg~s$^{-1}$ ($L_{\rm
X}<3.0\times10^{38}$~erg~s$^{-1}$).

In addition to these global properties, the {\it XMM-Newton}
observation provides constraints on the spatial, spectral, and temporal
properties of the X-ray point-source population as detailed below.

\subsection{Spatial Extent of the Nuclear Source XMMU~J034648.8$+$680546 (IC~342~X-3)}\label{X3_spatial}

\begin{figure}
\vspace{-0.1in}
\centerline{
\includegraphics[width=8.5cm, angle=0]{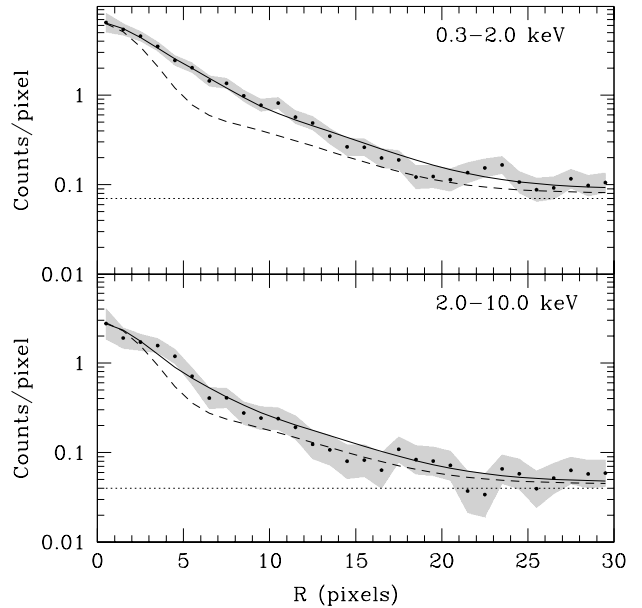}
}
\vspace{-0.02in} 
\figcaption[IC342.eps]{
The soft and hard-band MOS1$+$MOS2 radial profiles (small filled
circles) of the IC~342 nucleus, compared to our simple model (solid
curves) and the {\it XMM--Newton} PSF (dashed curves) calculated at
1.5~keV and 6~keV, respectively; see $\S$\ref{X3_spatial} for
details. The average background has been added to both the model and
the PSF. The shaded regions indicate the 1$\sigma$ deviation of the
measured profile. Deviations from the PSF, particularly in the soft
band, can be seen out to radii of $\approx$20 and $\approx$10 pixels,
at which point the 0.07~counts~pixel$^{-1}$ soft-band and
0.04~counts~pixel$^{-1}$ hard-band backgrounds begin to dominate
(dotted lines), respectively. A MOS pixel is equivalent to 1\farcs1.
\label{fig:IC342-nucleus-profile}}
\vspace{-0.1in} 
\end{figure}

This {\it XMM-Newton} observation offers an order of magnitude
improvement compared to the high-resolution {\it ROSAT} HRI
observations in the number of counts collected from the nucleus of
IC~342, allowing us to improve measurements of the nuclear spatial
extent reported by BCT93. The nuclear source appears to be azimuthally
symmetric in the raw images out to $\sim$40$\arcsec$, so we extracted
soft and hard-band counts from it in 1\farcs1 (1 MOS pixel) annular
bins using the combined MOS1$+$MOS2 image; the spatial resolution of
the PN is $\ga1.5\times$ worse ({\it XMM-Newton} Users' Handbook
$\S$3.2.1.1) and is therefore not very useful for this purpose. We
also generated images of the MOS1$+$MOS2 PSF using the {\tt SAS}
calibration tool {\tt calview}. Counts from these images were then
extracted in an identical manner to the data and normalized by the
innermost nuclear annulus. The resulting radial surface-brightness
distributions and 1$\sigma$ errors for the data are shown in
Figure~\ref{fig:IC342-nucleus-profile}.

The surface-brightness distribution of the nucleus lies above the
background out to radii of $\sim$15--20 pixels
\hbox{($\sim16$--$22\arcsec$)} and is clearly extended compared to the
PSF model. The accuracy of the MOS PSF model from {\tt calview} was
verified using source 26 ($1\arcmin$ from the nucleus); deviations
between this source and the PSF were much smaller than those seen in
Figure~\ref{fig:IC342-nucleus-profile} and were consistent with source
statistics. The fact that the fraction of extended emission is larger
in the soft band, even though the imaging quality is similar in both
bands, further supports the reality of the extent. An emission model
consisting of both a point source and a uniform disk with an
$\approx8\arcsec$ (128~pc) radius convolved with the instrument PSF
represents the surface-brightness distribution well, although there is
clearly still some residual scatter due to possible clumping or patchy
absorption. This is especially noticeable in the hard band where
individual point sources are more likely to dominate the emission. The
relative contributions from the extended disk component are
$\approx55$\% and $\approx35$\% in the soft and hard bands,
respectively, in crude agreement with the BCT93 soft-band values
($9\arcsec$ radius, $44$\%). The relative softness of the extended
emission is consistent with its expected physical origin as
supernova-heated hot gas (see also $\S$\ref{X3} for X-ray spectral
details). Moreover, the extent of the X-ray disk emission mirrors the
surface brightness enhancements seen at optical-to-radio
wavelengths \citep[\hbox{$\approx$~10--15\arcsec}; e.g.,][]{Boker1997,
Buta1999, Schinnerer2003} that are attributed to a starburst ring,
further strengthening our conclusions.

\subsection{X-ray Timing Analysis}\label{timing}

\subsubsection{Short-Term Variability}\label{short-term}

Although the {\it XMM-Newton} observation of IC~342 was short, we do
have adequate statistics to evaluate the short-term timing
characteristics of some of the off-nuclear sources. To examine
objectively the existence of any significant variations in the count
rate, we used the Kolmogorov-Smirnov (KS) statistic on the unbinned
data to test the null hypothesis that each source plus background rate
was constant over the duration of the exposure. None of the sources
varied at the $>$~90\% confidence level.

\subsubsection{Long-Term Variability}\label{long-term}

Comparison of the {\it XMM-Newton} observation with the 19~ks {\it
ROSAT} HRI observation on 1991 February 13--16 allows us to
constrain the soft-band long-term variability of the entire IC~342
X-ray source population, while comparison with the 38~ks {\it ASCA}
observation on 1993 September 19 and the 276~ks {\it ASCA} observation
on 2000 February 24--March 1 allows us to constrain the full-band
variability for X-1, X-2, and X-3.  The hard-band variability will
be discussed in $\S$\ref{spectra} in the context of our X-ray spectral
analyses.

To measure the soft-band variability, we extracted the HRI counts at
the position of each source using the {\tt XIMAGE} software package,
which corrects for vignetting and aperture losses. For detections, we
used the values reported in BCT93, and for 3$\sigma$ upper limits we
used the {\it sosta} algorithm. The HRI count rates were converted to
0.5--2~keV fluxes using {\sc pimms} (column~10 of
Table~\ref{tab:sources}). The {\sc pimms} input spectral model was set
to the best-fit model found from our {\tt XSPEC} spectral analyses in
$\S$\ref{spectra}. We found that five sources were variable in the
soft band over the 10-year period: XMMU~J034556.0$+$680455 and
XMMU~J034646.0$+$680946 faded by a factor of $\approx$6 and
$\approx$3.3. compared to the {\it ROSAT} observation, respectively;
XMMU~J034719.2$+$681129 and XMMU~J034723.3$+$680857 brightened by at
least factors of 1.5 and 1.4 compared to their {\it ROSAT} upper
limits, respectively; and finally J034541.4$+$680241 ({\it ROSAT}
source 2; BCT93) was not detected by our {\it XMM-Newton} observation
at all, indicating a decrease of at least a factor of 5. Given the
poor HRI statistics for this latter source, however, it may have been
spurious, although strong variability or extreme X-ray spectral
properties (i.e., a very soft spectrum) cannot be ruled out.

\tabletypesize{\scriptsize}
\begin{deluxetable*}{rcrrcl}
\tablecaption{X-ray Spectral Analysis Results\label{tab:spectra}} 
\tablehead{
\colhead{(1)} & 
\colhead{(2)} & 
\colhead{(3)} & 
\colhead{(4)} & 
\colhead{(5)} & 
\colhead{(6)} \\
\colhead{ID} & 
\colhead{Best-fit Model} & 
\colhead{$N_{\rm H}$} & 
\colhead{$\Gamma/kT/T_{\rm in}$} & 
\colhead{$\chi^2$/DOF} & 
\colhead{Comments}
}
\startdata
 7         & P      & $6.0^{+0.5}_{-0.5}$  & $1.68^{+0.09}_{-0.08}$                        &  166.0/193          & BB model also acceptable, but shows systematic residuals\\
13         & BB     & $15.3^{+1.9}_{-1.8}$ & $2.17^{+0.25}_{-0.12}$                        &  113.0/108          & P model shows somewhat worse residuals \\
19         & P      & $4.6^{+1.8}_{-1.4}$  & $1.84^{+0.39}_{-0.16}$                        &   17.6/16           & \\
21         & T$+$P  & $6.4^{+0.7}_{-1.0}$  & $kT=0.30^{+0.33}_{-0.07}$/$\Gamma=2.52^{+0.15}_{-0.18}$ &   94.4/97 & One-component models are not acceptable, $Z=6.39(> 1.45) Z_{\odot}$ \\
26         & BB     & $2.0^{+0.9}_{-0.7}$  & $1.46^{+0.21}_{-0.17}$                        &   45.1/52           & P and T models also acceptable, but show some systematic residuals\\
\hline \\
\vspace{0.1cm} 
ID         & Best-fit Model & $N_{\rm H}$ & $\Gamma/kT/T_{\rm in}$       &   $cstat$/DOF          & Comments\\
\hline \\
 6         & P      & $6.7^{+1.1}_{-2.5}$  & $1.77^{+0.16}_{-0.16}$                        & 1375.5/3233$\dagger$ & T model also acceptable, but shows worse residuals \\
 9         & P      & $5.2^{+0.8}_{-1.0}$  & $1.82^{+0.16}_{-0.14}$                        & 1255.7/3233$\dagger$ & \\
11{\rlap*} & T      & $6.2^{+0.7}_{-0.6}$  & $0.17^{+0.03}_{-0.01}$                        &  870.6/3233$\dagger$ & Soft X-ray spectrum, consistent with star ID \\
20{\rlap*} & T      & $2.8^{+3.6}_{-1.9}$  & $0.25^{+0.04}_{-0.08}$                        &  706.2/3233$\dagger$ & Soft X-ray spectrum, consistent with faint star ID, $Z=0.28^{+0.38}_{-0.13}Z_{\odot}$ \\
30         & P      & $6.9^{+4.3}_{-1.6}$  & $3.14^{+1.54}_{-0.18}$                        &  842.8/3233$\dagger$ & T and BB models also acceptable. \\
31         & P      & $8.6^{+0.8}_{-2.7}$  & $3.79^{+0.41}_{-0.76}$                        &  969.8/3233$\dagger$ & T and BB models also acceptable. \\
33         & P      & $4.7^{+0.8}_{-0.7}$  & $2.17^{+0.19}_{-0.18}$                        & 1046.1/3233$\dagger$ & \\
\enddata
\tablecomments{
Col. 1: ID number. Sources denoted by a ``*'' are not thought to be intrinsic to IC~342.
Column 2: Best-fit spectral model from XSPEC for bright sources ($>$
150 counts). For sources with fewer than 400 total counts, the Cash
statistic \citep{Cash1979} was used instead of $\chi^2$. All models
were fit with a {\it wabs} absorption component. A ``P'' indicates a
power-law component, a ``T'' indicates a {\it mekal} thermal-plasma
component, and a ``BB'' indicates a multi-color disk component. Note
that the Galactic column toward IC~342 is $N_{\rm H} =
3\times10^{21}$~cm$^{-2}$
\citep{Stark1992}.
Column 3: Neutral hydrogen absorption column density in units of
$10^{21}$ cm$^{-2}$ as determined from the best-fit models to the {\it
XMM-Newton} spectra. Also listed are the 90\% confidence errors
calculated for one parameter of interest ($\Delta\chi^2 = 2.7$).
Column 4: Power-law photon index $\Gamma$, thermal-plasma temperature
$kT$ (keV), or multi-color disk inner temperature $T_{\rm in}$ as
determined from the best-fit models to the {\it XMM-Newton}
spectra. Also listed are the 90\% confidence errors calculated for one
parameter of interest ($\Delta\chi^2 = 2.7$).
Column 5: $\chi^2$ and degrees of freedom. Sources denoted by a
``$\dagger$'' were fit with the Cash statistic on the unbinned data,
in which case the ``cstat'' value and degrees of freedom are listed.
Column 6: Comments. 
}
\end{deluxetable*}

\subsection{X-ray Spectral Analysis}\label{spectra}

For each source, we extracted the spectra and response matrices for
each CCD camera separately and fit all three spectra with the same
model in {\tt XSPEC}. We initially fitted the spectra with absorbed
power-law models.\footnote{In most cases, a {\it mekal} thermal plasma
model \citep{Mewe1985} was equally acceptable.} The spectral
properties of the 23 sources with fewer than 150 full-band counts are
not well constrained and have been fitted only with a mean spectrum as
given in $\S$\ref{data} to derive X-ray fluxes and luminosities. The
12 sources with more than 150 full-band counts all had adequate photon
statistics for individual fits, although $\ga1000$ counts were
typically necessary to rule out competing spectral models clearly. The
best-fit spectral parameters for these 12 sources are given in
Table~\ref{tab:spectra}.  All of the sources thought to be intrinsic
to IC~342 exhibited spectral cutoffs below $\sim$~1~keV that are
best fit with column densities equal to or larger than the Galactic
value (see $\S$\ref{intro}), consistent with these off-nuclear X-ray
sources being located in star-forming regions and spiral arms within
IC~342. We discuss our results for the four brightest X-ray sources in
IC~342 with $>1000$ counts below.

\begin{figure*}
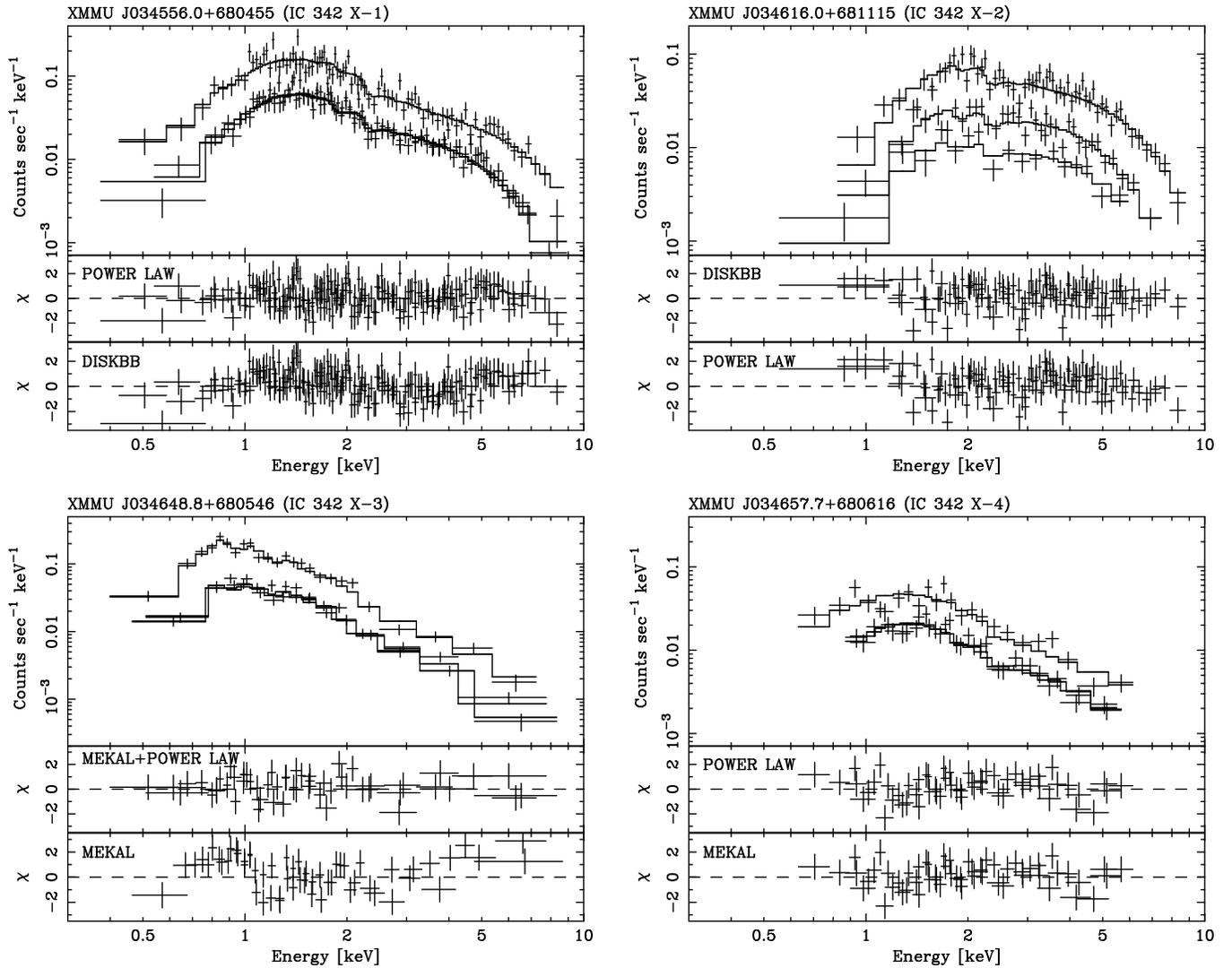

\vspace{0.1in}
\centerline{
\includegraphics[width=7cm, angle=-90]{fig4a.ps}
\hfill
\includegraphics[width=7cm, angle=-90]{fig4b.ps}
}
\vspace{0.1in}
\centerline{
\includegraphics[width=7cm, angle=-90]{fig4c.ps}
\hfill
\includegraphics[width=7cm, angle=-90]{fig4d.ps}
}
\vspace{0.1cm} 
\figcaption[IC342.eps]{
X-ray spectra of the four brightest X-ray sources in IC~342. The upper
panel of each plot shows the PN, MOS1, and MOS2 spectra and best-fit
models (see $\S$\ref{spectra} for details). The lower two panels show
the residuals of the fits measured in units of $\sigma$ for the
best-fit model (middle) and the next most likely model (bottom).
\label{fig:spectra}}
\vspace{0.2cm} 
\end{figure*}

\subsubsection{XMMU~J034556.0$+$680455 (IC~342 X-1)}\label{X1}

IC~342 X-1 is the brightest source in IC~342 and was recently found to
be coincident with a supernova remnant \citep{Roberts2003}.  X-1 has
been extensively studied by {\it ASCA}, as it is considered to be one
of the best examples of a massive ($\ga10M_{\odot}$) black-hole
binary. O98 found this source in a luminous high state in 1993, with a
spectrum best-fitted by an absorbed multi-color disk model \citep[{\it
diskbb} model in {\sc xspec};][]{Mitsuda1984} arising from an
optically thick standard accretion disk around a black hole. The
best-fit absorbed {\it diskbb} parameters were $N_{\rm
H}=(4.7\pm0.3)\times10^{21}$~cm$^{-2}$ and $T_{\rm
in}=1.77\pm0.05$~keV, resulting in a 0.5--10~keV flux of $F_{\rm
0.5-10~keV}\approx 1.0\times10^{-11}$~erg~cm$^{-2}$~s$^{-1}$. When it
was revisited in 2000, K01 found the flux ($F_{\rm 0.5-10~keV}\approx
3.1\times10^{-12}$~erg~cm$^{-2}$~s$^{-1}$) and spectrum of this source
had changed dramatically, and it was best fit by an absorbed power-law
model with $N_{\rm H}=(6.4\pm0.7)\times10^{21}$~cm$^{-2}$ and $\Gamma
= 1.73\pm0.06$.  An ionized Fe-K edge at $8.4\pm0.3$~keV with an
optical depth of $0.9\pm0.5$ was also seen. K01 attributed this
spectral change to be a transition between a high/soft state and a
low/hard state, as is observed in many Galactic and Magellanic
black-hole binary systems. \citet[][hereafter K02]{Kubota2002}
re-examined the 2000 {\it ASCA} data, finding that the spectrum
deviates from a power-law shape such that there appears to be
significant softening above 5~keV. They proposed that the spectrum was
not indicative of a low-hard state, but rather of an anomalous very
high (VH) state as observed in some Galactic black-hole binaries,
which was adequately fit by a strongly Comptonized optically thick
accretion disk with $T_{\rm in}=1.1\pm0.3$~keV and $\Gamma^{\rm
th}=2.2\pm0.4$. We note that the {\it ASCA} $\approx$2\farcm5 radius
beam used to study X-1 (compare with Figure~1 of O98) would have also
included sources 6, 9, and 11 from Table~\ref{tab:sources},
contaminating the {\it ASCA} source by $\approx19$\% based on the
relative {\it XMM--Newton} source fluxes.

To compare with the {\it ASCA} spectral fits, we fitted both an
absorbed {\it diskbb} model and an absorbed power law to the
X-ray spectrum. The best-fit parameters for the absorbed {\it diskbb}
model were $T_{\rm in}=1.93^{+0.14}_{-0.13}$~keV and $N_{\rm
H}=(3.5^{+0.3}_{-0.3})\times10^{21}$~cm$^{-2}$ ($\chi^2=206.9$ for 193
degrees of freedom), while those for the absorbed power law were
$\Gamma= 1.68^{+0.09}_{-0.08}$ and $N_{\rm
H}=(6.0^{+0.5}_{-0.5})\times10^{21}$~cm$^{-2}$ ($\chi^2=166.0$ for 193
degrees of freedom). Although the {\it diskbb} fit is formally
acceptable, there are systematic residuals at the softest and hardest
energies, and hence the power-law model appears to offer a superior
fit to the data (see the X-1 residuals in
Figure~\ref{fig:spectra}). The flux of the source is found to be
$F_{\rm 0.5-10~keV}\approx 2.15\times10^{-12}$~erg~cm$^{-2}$~s$^{-1}$
The best-fit parameters from the power-law model are consistent with
those obtained during the 2000 {\it ASCA} observation, suggesting that
little changed spectrally in the $\approx$27.5 months between the
latest {\it ASCA} observation and our {\it XMM-Newton}
observation. The {\it XMM-Newton} flux, on the other hand, is
$\approx$~30\% lower than the 2000 {\it ASCA} measurement. This is a
bit larger than the source confusion estimate made above (accounting
for slight bandpass differences), but the difference is only
marginally significant. Thus we conclude that X-1 has remained
relatively constant over the $\approx 2$~yrs between the latest {\it
ASCA} and {\it XMM--Newton} observations and is therefore likely to be
in a VH state similar to the one reported by K02.

If X-1 truly is in a VH state, and this state has remained roughly
constant for $\approx 2$~yrs, then it marks the longest period over
which a VH state has been observed. For comparison, this would be
$\approx10\times$ longer than has been observed in GX~339-4, a
Galactic binary which went into a VH state for $\approx$3~months
(M. Nowak 2003, private communication). Given the limited amount of
information regarding the VH state in general, the longevity of the VH
state observed in X-1 is plausibly consistent with current
understanding of this state.  Another possibility is that both the
2000 {\it ASCA} and 2002 {\it XMM-Newton} observations caught X-1 in a
VH state, but it varied out of the VH state in
between. GRS~1915$+$105, for instance, is known to vary in and out of
the VH state frequently and has a non-negligible duty cycle in
the VH state \citep[e.g.,][]{Belloni2000}. 

\subsubsection{XMMU~J034616.0$+$681115 (IC~342 X-2)}\label{X2}

IC~342~X-2 is the second brightest source in IC~342 and also has been
extensively studied by {\it ASCA}. In contrast to X-1, this source was
in a low state in 1993 with $F_{\rm 0.5-10~keV}\approx
4.1\times10^{-12}$~erg~cm$^{-2}$~s$^{-1}$. Its X-ray spectrum was
acceptably fit by either an absorbed power-law or {\it diskbb} model (O98,
K01). The best-fit absorbed power-law parameters were $N_{\rm
H}=(14.3\pm1.6)\times10^{21}$~cm$^{-2}$ and a photon index of
$\Gamma=1.39\pm0.10$~keV. In 2000, K01 found that the flux of this
source had nearly doubled to $F_{\rm 0.5-10~keV}\approx
7.2\times10^{-12}$~erg~cm$^{-2}$~s$^{-1}$, and its spectrum was best
fit by an absorbed {\it diskbb} model with $N_{\rm
H}=(18.0\pm8.0)\times10^{21}$~cm$^{-2}$ and $T_{\rm
in}=1.62\pm0.04$~keV. K01 attributed this to be a spectral transition
between a low/hard state and a high/soft state.  As with X-1, we note
that the {\it ASCA} $\approx$3\farcm0 radius beam used to study X-2
would have also included source 12 and possibly source 20 (compare
with Figure~1b of K01). The contamination from these sources to the
{\it ASCA} source, however, is only $\approx3$--10\% based on the
relative {\it XMM--Newton} source fluxes (and should be even less
after the strong variability seen from X-2 is accounted for; see
below).

Based on the {\it ASCA} spectral fits, we fitted both absorbed {\it
diskbb} and power-law models to the X-ray spectrum of X-2. We note
that X-2 lies along a chip gap in the MOS2 detector, resulting in a
50\% loss of counts with that device. The best-fit parameters for the
absorbed {\it diskbb} model were $T_{\rm in}=2.17^{+0.25}_{-0.12}$~keV
and $N_{\rm H}=(15.3^{+1.9}_{-1.8})\times10^{21}$~cm$^{-2}$
($\chi^2=113.0$ for 108 degrees of freedom), while those for the
absorbed power-law model were $\Gamma=1.81\pm0.18$ and $N_{\rm
H}=(22.5^{+3.3}_{-2.9})\times10^{21}$~cm$^{-2}$ ($\chi^2=127.3$ for
108 degrees of freedom). Both models provide acceptable fits to the
data, although the residuals of the {\it diskbb} model exhibit fewer
systematic deviations (see Figure~\ref{fig:spectra}). The best-fit
parameters are different from those obtained during either {\it ASCA}
observation, demonstrating that X-2 has undergone further spectral
variability. In addition, the {\it XMM-Newton} flux has dropped by
factors of $\sim$1.7 and $\sim$3 compared with the {\it ASCA} results
of 1993 and 2000. Thus although this state is spectrally intermediate
between the 1993 and 2000 states, it is the lowest flux state
exhibited by X-2 to date. It is interesting that this new low/hard
state is better fit by the {\it diskbb} model than a power-law model,
as this indicates that X-2 still has some spectral curvature. This
curvature is in the opposite sense to that expected from a soft
thermal component with a hard tail (i.e., the standard for X-ray
binaries in their high state). The column density observed toward this
source, however, is a factor of $\approx$2--3 higher than seen from
any of the other sources, suggesting that X-2 has significant local
absorption. It is therefore possible that the apparent curvature is
due to complex absorption near the source, which is also often
detected in luminous Galactic X-ray binaries.

\subsubsection{XMMU~J034648.8$+$680546 (IC~342~X-3)}\label{X3}

IC~342~X-3 is coincident with the nucleus and was clearly detected by
both the {\it ROSAT} HRI and {\it ASCA}. The spectral properties of
X-3 from the 1993 {\it ASCA} observation are discussed briefly in O98,
who found that the source was best fitted by an absorbed power law
with $\Gamma=2.1\pm0.2$ and fixed $N_{\rm
H}=3.0\times10^{21}$~cm$^{-2}$. Again, we note that the {\it ASCA}
$\approx$3\farcm0 radius beam used to study X-3 would have also
included sources 19, 24, 26, and 29 (compare with Figure~1 of
O98). The contamination from these sources to the {\it ASCA} source is
$\approx44$\% based on the relative {\it XMM--Newton} source fluxes.

We initially fitted both single absorbed power-law and thermal-plasma
models to the X-ray spectrum of the nucleus. Both fits were
unacceptable, with the absorbed power-law model giving $\Gamma=2.67$
and $N_{\rm H}=4.0\times10^{21}$~cm$^{-2}$ ($\chi^2=180.4$ for 101
degrees of freedom), and the absorbed {\it mekal} model giving
$kT=2.14$~keV, $Z=0.02Z_{\odot}$, and $N_{\rm
H}=2.4\times10^{21}$~cm$^{-2}$ ($\chi^2=214.9$ for 100 degrees of
freedom). Neither model acceptably fits the continuum shape or the
apparent Fe L-shell emission complexes visible at $\approx$0.8~keV and
$\approx$1.0~keV (see Figure~\ref{fig:spectra}). The presence of Fe
L-shell emission at these two energies suggests that a large fraction
of the emission is likely to be thermal in origin. In
$\S$\ref{X3_spatial} we were able to separate spatially X-3 into
point-like and extended components. Thus we should expect that the
spectrum of X-3 might be complex, consisting of perhaps two or more
components. We therefore modeled the spectrum with an absorbed {\it
mekal}$+$power-law model. This model provides a significant
improvement over the single-component models ($\chi^2=94.4$ for 96
degrees of freedom), with best-fit values of
$kT=0.30^{+0.33}_{-0.07}$~keV, $\Gamma=2.52^{+0.15}_{-0.18}$, $Z=6.39
(> 1.45) Z_{\odot}$, and $N_{\rm
H}=(6.8^{+1.6}_{-1.5})\times10^{21}$~cm$^{-2}$. An absorbed {\it
mekal}$+${\it mekal} model with a single metallicity also provides an
acceptable fit to the data ($\chi^2=97.9$ for 96 degrees of freedom),
with best-fit values of $kT_{1}=0.50^{+0.12}_{-0.22}$~keV,
$kT_{2}=4.02^{+1.89}_{-1.30}$~keV, $Z=0.12^{+0.08}_{-0.07}Z_{\odot}$,
and $N_{\rm H}=(5.0^{+2.2}_{-1.1})\times10^{21}$~cm$^{-2}$. However,
the best-fit abundance for this second model is low compared to the
optically derived value of $\approx 3 Z_{\odot}$
\citep{Diaz1986,Vila-Costas1992}. Fixing the absorbed {\it
mekal}$+${\it mekal} model abundance at $3 Z_{\odot}$ (or even
$Z_{\odot}$) results in an unacceptable fit with $\chi^2=169.9$
($\chi^2=120.5$) for 97 degrees of freedom.  The absorbed {\it
mekal}$+$power-law model on the other hand provides a consistent
measure of the nuclear abundance.

The relative contribution of the mekal component to the total absorbed
flux of X-3 is 45\% and 1\% in the soft and hard bands,
respectively. The soft-band fraction is in basic agreement with the
physical picture found from our spatial analysis
($\S$\ref{X3_spatial}), suggesting that the soft thermal plasma
component is likely to be spatially extended and of a starburst
origin. The complete domination of the power-law component above
2~keV, coupled with the fact that we still find significant amount
extended emission in this band, suggests that the hard component is
probably complex and may be due to an unknown number of circumnuclear
X-ray binaries.\footnote{Another possible origin for the point-like
fraction of the point-like emission is from a low-luminosity active
nucleus, although there is no evidence for one from past studies of
the nuclear region.} The absorption-corrected 0.5--10.0~keV luminosity
of the extended nuclear component is $L_{\rm
X}\approx2.1\times10^{39}$~erg~s$^{-1}$), which is comparable to or
larger than our constraint on extended emission from the rest of
IC~342.

\subsubsection{XMMU~J034657.7$+$680616 (IC~342~X-4)}\label{X4}

IC~342~X-4 is a bright source located $\approx$1\farcm0 from the
nucleus, which was only resolved by the {\it ROSAT} HRI observation
({\it ROSAT} source 9; BCT93). The source was presumably detected by
{\it ASCA} as well but was blended together with the nucleus in those
observations, contributing $\approx31$\% based on the relative {\it
XMM--Newton} source fluxes. We initially fitted X-4 with an absorbed
power-law model, obtaining an acceptable fit with best-fit parameters
of $N_{\rm H}=(4.4^{+1.4}_{-1.0})\times10^{21}$~cm$^{-2}$ and
$\Gamma=1.80^{+0.20}_{-0.09}$ ($\chi^2=51.8$ for 52 degrees of
freedom). We found that {\it diskbb} and {\it mekal} models provided
equally acceptable fits. However, while the {\it diskbb} model fit the
continuum shape best (see the X-4 residuals in
Figure~\ref{fig:spectra}), the best-fit absorption column value lies
just below Galactic. In the case of the {\it mekal} model, the high
temperature would imply that X-4 must be a young supernova, although
it clearly lacks the long-term variability or high equivalent-width
X-ray emission lines characteristic of such sources.  Thus it seems
that the power-law model is the most physically appropriate model. If
X-4 is a single X-ray binary, then its low photon index may indicate
that it is in a low/hard state; however, its spectral ambiguity leaves
room for several other possibilities as well.

\section{Conclusions}\label{conclusions}
Our {\it XMM--Newton} observation has tripled the number of known
X-ray point sources in IC~342 and has provided a number of new
constraints on the X-ray properties of both the point-source
population and any potential diffuse emission.

\begin{itemize}
\item We find a relatively flat XLF slope of 0.5$\pm$0.1 for IC~342, which 
is comparable to those found for actively star-forming galaxies such
as M82 and the Antennae and is inconsistent with those of quiescent
spiral galaxies \citep[e.g.,][]{Kilgard2002, Colbert2003}. This
implies that IC~342 may have played host to a significant amount
($\ga0.5$~$M_{\odot}$~yr$^{-1}$) of recent star-formation activity.
However, our upper limit on diffuse X-ray emission associated with hot
gas ($L_{\rm X}<5.6\times10^{39}$~erg~s$^{-1}$ for a 0.5~keV thermal
plasma), and its infrared luminosity of $L_{60~\mu
m}=9\times10^{42}$~erg~s$^{-1}$ \citep{Shapley2001}, are more typical
of other Sab galaxies such as M58 or M94
\citep[e.g.,][]{Eracleous2002} rather than luminous starbursts
\citep[e.g.,][]{Read1997}. Thus IC~342 appears to have the X-ray binary
population of a starburst, but lacks the hot gas and luminous infrared
emission that typically accompanies vigorous star formation. This
suggests that the mode of star formation in IC~342 is probably quite
different from that of archetypal starburst sources.

\item We find no evidence for short-term variability among the X-ray 
sources in this short observation. We do find long-term variability
between this observation and the 1991 {\it ROSAT} and 1993/2000 {\it
ASCA} observations for five sources.

\item We find X-1 to be in a state identical to the one observed 
by {\it ASCA} in 2000, after correcting the {\it ASCA} measurement for
source contamination. If X-1 is in a VH state as contended by K02, and
it has remained in this state between 2000 and 2002, then it would be
the longest duration VH-state binary ever observed. Alternatively, it
might be possible that both the 2000 {\it ASCA} and 2002 {\it
XMM-Newton} observations caught X-1 in a VH state by chance, but that
it has varied out of the VH state in between.

\item We find X-2 has decreased in flux by factors of $\sim$1.7 and 
$\sim$3 compared to its states observed by {\it ASCA} in 1993 and
2000, respectively, while its spectrum is intermediate between the
low/hard 1993 state and the high/soft 2000 state. This is the lowest
luminosity state observed for X-2 to date.

\item We have confirmed that a large fraction ($\approx$55\% and 
$\approx$35\% in the 0.3--2.0~keV and 2.0--10.0~keV bands,
respectively) of the X-ray emission coincident with nucleus of IC~342
is associated with an extended, relatively uniform disk approximately
$8\arcsec$ in radius. The spectrum of the nucleus is best fit by an
absorbed two-component model consisting of a thermal plasma with a
temperature of $kT\approx0.3$ and a power law with a photon index of
$\Gamma\approx2.53$. The relative fluxes of the two spectral
components suggest that the nucleus is complex, with a soft extended
component contributing $\approx$50\% of the total luminosity.
\end{itemize}

Because of their relative distance and luminosity, the X-ray binaries
in IC~342 provide an excellent laboratory to advance our knowledge of
ultraluminous X-ray sources even further. For instance, a higher
signal-to-noise spectrum of X-1 with {\it XMM-Newton} or {\it
Constellation-X} might provide stronger constraints on its nature,
especially regarding the origin of the spectral softening above 4~keV
and an apparent spectrum bump around 1.5~keV. Uniform temporal
monitoring of X-1, X-2, and X-3 could provide more information about
the frequency and amplitude of spectral transitions, which would help
place stronger limits on their black hole masses. Finally, higher
resolution X-ray imaging of the nucleus with {\it Chandra} would allow
a decoupling of the extended and point-like emission with the two
spectral components.

\acknowledgements
We thank I.~Lehmann, M.~Nowak, and C.~Vignali for helpful discussions,
and R.~Buta and A.~Saha for use of their optical images.
%
%
We gratefully acknowledge the financial support of NASA awards
NAG5-10089 (FEB), NAG5-9940 (BL, WNB), and LTSA NAG5-13035.


\begin{thebibliography}{34}
\expandafter\ifx\csname natexlab\endcsname\relax\def\natexlab#1{#1}\fi
\expandafter\ifx\csname url\endcsname\relax
  \def\url#1{{\tt #1}}\fi
\expandafter\ifx\csname urlprefix\endcsname\relax\def\urlprefix{URL }\fi
\providecommand{\eprint}[2][]{\url{#2}}

\bibitem[\protect\astroncite{{Arnaud}}{1996}]{Arnaud1996}
{Arnaud}, K.~A. 1996, in ASP Conf. Ser. 101: Astronomical Data Analysis
  Software and Systems V, vol.~5, 17--20

\bibitem[\protect\astroncite{{Baldi} et~al.}{2002}]{Baldi2002}
{Baldi}, A., {Molendi}, S., {Comastri}, A., {Fiore}, F., {Matt}, G., \&
  {Vignali}, C. 2002, \apj, 564, 190

\bibitem[\protect\astroncite{{Belloni} et~al.}{2000}]{Belloni2000}
{Belloni}, T., {Klein-Wolt}, M., {M{\' e}ndez}, M., {van der Klis}, M., \& {van
  Paradijs}, J. 2000, \aap, 355, 271

\bibitem[\protect\astroncite{{B\"{o}ker} et~al.}{1997}]{Boker1997}
{B\"{o}ker}, T., {Forster-Schreiber}, N.~M., \& {Genzel}, R. 1997, \aj, 114,
  1883

\bibitem[\protect\astroncite{{Bregman} et~al.}{1993}]{Bregman1993}
{Bregman}, J.~N., {Cox}, C.~V., \& {Tomisaka}, K. 1993, \apjl, 415, L79,
  (BCT93)

\bibitem[\protect\astroncite{{Buta} \& {McCall}}{1999}]{Buta1999}
{Buta}, R.~J. \& {McCall}, M.~L. 1999, \apjs, 124, 33

\bibitem[\protect\astroncite{{Cash}}{1979}]{Cash1979}
{Cash}, W. 1979, \apj, 228, 939

\bibitem[\protect\astroncite{{Colbert} et~al.}{2003}]{Colbert2003}
{Colbert}, E., {Heckman}, T., {Ptak}, A., \& {Strickland}, D. 2003, \apj,
  submitted (astro-ph/0305476)

\bibitem[\protect\astroncite{{Crosthwaite} et~al.}{2000}]{Crosthwaite2000}
{Crosthwaite}, L.~P., {Turner}, J.~L., \& {Ho}, P.~T.~P. 2000, \aj, 119, 1720

\bibitem[\protect\astroncite{{Diaz} \& {Tosi}}{1986}]{Diaz1986}
{Diaz}, A.~I. \& {Tosi}, M. 1986, \aap, 158, 60

\bibitem[\protect\astroncite{{Ebeling} et~al.}{2003}]{Ebeling2003}
{Ebeling}, H., {White}, D., \& {Rangarajan}, F.~V.~N. 2003, \mnras, 0,
  submitted

\bibitem[\protect\astroncite{{Eracleous} et~al.}{2002}]{Eracleous2002} 
Eracleous, M., Shields, J.~C., Chartas, G., \& Moran, E.~C.\ 2002, \apj, 
565, 108 

\bibitem[\protect\astroncite{{Freeman} et~al.}{2002}]{Freeman2002}
{Freeman}, P.~E., {Kashyap}, V., {Rosner}, R., \& {Lamb}, D.~Q. 2002, \apjs,
  138, 185

\bibitem[\protect\astroncite{{Gehrels}}{1986}]{Gehrels1986}
{Gehrels}, N. 1986, \apj, 303, 336

\bibitem[\protect\astroncite{{Jansen} et~al.}{2001}]{Jansen2001}
{Jansen}, F., et~al. 2001, \aap, 365, L1

\bibitem[\protect\astroncite{{Kilgard} et~al.}{2002}]{Kilgard2002}
{Kilgard}, R.~E., {Kaaret}, P., {Krauss}, M.~I., {Prestwich}, A.~H., {Raley},
  M.~T., \& {Zezas}, A. 2002, \apj, 573, 138

\bibitem[\protect\astroncite{{Kubota} et~al.}{2002}]{Kubota2002}
{Kubota}, A., {Done}, C., \& {Makishima}, K. 2002, \mnras, 337, L11, (K01)

\bibitem[\protect\astroncite{{Kubota} et~al.}{2001}]{Kubota2001}
{Kubota}, A., {Mizuno}, T., {Makishima}, K., {Fukazawa}, Y., {Kotoku}, J.,
  {Ohnishi}, T., \& {Tashiro}, M. 2001, \apjl, 547, L119, (K01)

\bibitem[\protect\astroncite{{Lyons}}{1991}]{Lyons1991}
{Lyons}, L. 1991, {Data Analysis for Physical Sciences} (Cambridge, Cambridge
  University Press)

\bibitem[\protect\astroncite{{Maccacaro} et~al.}{1988}]{Maccacaro1988}
{Maccacaro}, T., {Gioia}, I.~M., {Wolter}, A., {Zamorani}, G., \& {Stocke},
  J.~T. 1988, \apj, 326, 680

\bibitem[\protect\astroncite{{Mewe} et~al.}{1985}]{Mewe1985}
{Mewe}, R., {Gronenschild}, E.~H.~B.~M., \& {van den Oord}, G.~H.~J. 1985,
  \aaps, 62, 197

\bibitem[\protect\astroncite{{Mitsuda} et~al.}{1984}]{Mitsuda1984}
{Mitsuda}, K., et~al. 1984, \pasj, 36, 741

\bibitem[\protect\astroncite{{Monet} et~al.}{2003}]{Monet2003}
{Monet}, D.~G., et~al. 2003, \aj, 125, 984

\bibitem[\protect\astroncite{{Okada} et~al.}{1998}]{Okada1998}
{Okada}, K., {Dotani}, T., {Makishima}, K., {Mitsuda}, K., \& {Mihara}, T.
  1998, \pasj, 50, 25, (O98)

\bibitem[\protect\astroncite{{Read} et~al.}{1997}]{Read1997}
{Read}, A.~M., {Ponman}, T.~J., \& {Strickland}, D.~K. 1997, \mnras, 286, 626

\bibitem[\protect\astroncite{{Roberts} et~al.}{2003}]{Roberts2003}
{Roberts}, T.~P., {Goad}, M.~R., {Ward}, M.~J., \& {Warwick}, R.~S. 2003,
  \mnras, 342, 709

\bibitem[\protect\astroncite{{Saha} et~al.}{2002}]{Saha2002}
{Saha}, A., {Claver}, J., \& {Hoessel}, J.~G. 2002, \aj, 124, 839

\bibitem[\protect\astroncite{{Schinnerer} et~al.}{2003}]{Schinnerer2003}
{Schinnerer}, E., {B{\" o}ker}, T., \& {Meier}, D.~S. 2003, \apjl, 591, L115

\bibitem[\protect\astroncite{{Shapley} et~al.}{2001}]{Shapley2001}
{Shapley}, A., {Fabbiano}, G., \& {Eskridge}, P.~B. 2001, \apjs, 137, 139

\bibitem[\protect\astroncite{{Stark} et~al.}{1992}]{Stark1992}
{Stark}, A.~A., {Gammie}, C.~F., {Wilson}, R.~W., {Bally}, J., {Linke}, R.~A.,
  {Heiles}, C., \& {Hurwitz}, M. 1992, \apjs, 79, 77

\bibitem[\protect\astroncite{{Str{\" u}der} et~al.}{2001}]{Struder2001}
{Str{\" u}der}, L., et~al. 2001, \aap, 365, L18

\bibitem[\protect\astroncite{{Sugiho} et~al.}{2001}]{Sugiho2001}
{Sugiho}, M., {Kotoku}, J., {Makishima}, K., {Kubota}, A., {Mizuno}, T.,
  {Fukazawa}, Y., \& {Tashiro}, M. 2001, \apjl, 561, L73

\bibitem[\protect\astroncite{{Tully} \& {Fisher}}{1988}]{Tully1988}
{Tully}, R.~B. \& {Fisher}, J.~R. 1988, {Catalog of Nearby Galaxies} (Cambridge
  University Press, 1988)

\bibitem[\protect\astroncite{{Turner} et~al.}{2001}]{Turner2001}
{Turner}, M.~J.~L., et~al. 2001, \aap, 365, L27

\bibitem[\protect\astroncite{{Vila-Costas} \&
  {Edmunds}}{1992}]{Vila-Costas1992}
{Vila-Costas}, M.~B. \& {Edmunds}, M.~G. 1992, \mnras, 259, 121

\end{thebibliography}

\end{document}